\documentclass{iau307}
\usepackage{graphicx}
\usepackage{natbib}
\usepackage{url}
\usepackage{dtklogos}
\bibpunct{(}{)}{;}{a}{}{,}

\title[Fossil magnetic fields in rotating stars]
{Impact of rotation on the geometrical configurations of fossil magnetic fields}

\author[C. Emeriau \& S. Mathis]
{C. Emeriau$^{1}$
\and S. Mathis$^{1}$}

\affiliation{$^1$Laboratoire AIM Paris-Saclay, CEA/DSM - CNRS - Universit\'e Paris Diderot, IRFU/SAp Centre de Saclay, F-91191 Gif-sur-Yvette Cedex, France\\ email: {\tt constance.emeriau@cea.fr, stephane.mathis@cea.fr}}

\pubyear{2014}
\volume{307} 
\pagerange{}
\jname{New windows on massive stars: asteroseismology, interferometry, and spectropolarimetry}
\editors{G. Meynet, C. Georgy, J.H. Groh \& Ph. Stee, eds.}

\begin{document}

\maketitle

\begin{abstract}
The MiMeS project demonstrated that a small fraction of massive stars (around 7\%) presents large-scale, stable, generally dipolar magnetic fields at their surface. They are supposed to be fossil remnants of initial phases of stellar evolution. In fact, they result from the relaxation to MHD equilibrium states during the formation of stable radiation zones of initial fields generated by a previous convective phase. In contrast with the case of magnetic fields built by dynamo mechanisms, the geometry of fossil fields at the surface of early-type stars seems to be independent of rotation: dipolar fields are observed both in slowly- and rapidly-rotating stars. In this work, we present new theoretical results, where we generalized previous studies by taking rotation into account. The properties of relaxed fossil fields are compared to those obtained when rotation is ignored. Consequences for magnetic fields in the radiative envelope of rotating early-type stars are discussed.
\end{abstract}

\section{Fossil magnetic fields in early-type stars}

{Stellar magnetism is one of the key mechanisms that must be studied to understand the evolution of stars. Indeed, magnetic fields impact their structure, their activity and winds, and the transport of angular momentum and mixing in their interiors \citep[e.g.][Maeder, Eggenberger, and Gellert in this volume]{Mestel1999}. In this context, high-resolution spectropolarimetry allows us to study the large variety of stellar magnetic fields as a function of stellar parameters and evolutionary stages \citep[e.g.][Landstreet, and Wade in this volume]{DL2009}. In upper main-sequence stars with external radiative envelope, supposed {\it fossil} fields often appear as oblique dipoles at the surface \citep[][]{Wadeetal2011}. This dipolar geometry is observed both in slow and rapid rotators.}

\begin{figure}[h!]
\centering
\includegraphics[width=0.42\textwidth]{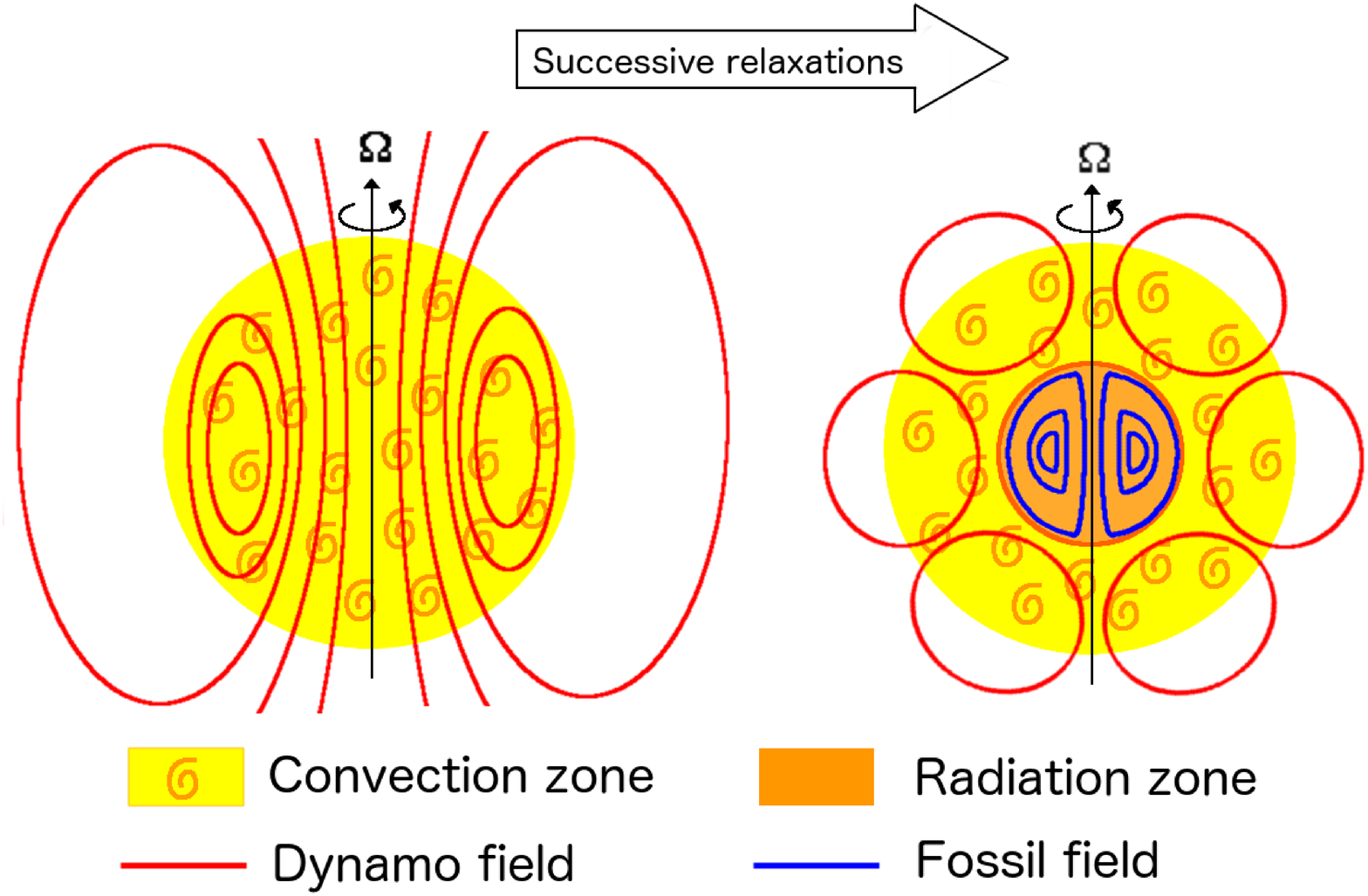} 
\caption{Scheme of the relaxation of fossil magnetic field at the time of the formation of an internal stably-stratified radiative zone during the PMS.}
\label{fig1}
\end{figure}

In this context, studies led by \cite{Tayler1973} and \cite{MarkeyTayler1973} demonstrated that purely poloidal and toroidal fields are unstable in stably stratified zones. It was then proposed by \cite{Tayler1980} that fossil fields have mixed configurations to be stable in such regions. This statement was verified by numerical simulations and theoretical works \citep[][]{BN2006,DBM2010} that found that in non-rotating experiments the lowest-energy equilibrium states are twisted non force-free dipolar fields. Moreover, fossil magnetic fields result from the adjustment of the initial field that has been generated by a previous convective phase (for example at the beginning of the Pre-Main-Sequence) during the transition from the turbulent convective state to a decaying turbulent one when the studied stable radiation zone is forming (see fig. \ref{fig1}, \cite{DM2010}, and the first numerical simulation of this mechanism during the recession of an external convective envelope computed by \cite{Arlt2014}). This process is known as \textit{turbulent relaxation} \citep[e.g.][]{Biskamp1997}.

\section{Modification of relaxed fossil magnetic fields by rotation}

In this work, we studied the impact of rotation on the properties of fossil magnetic fields by considering rotating magnetohydrodynamic relaxed equilibrium states using the method introduced by \cite{DM2010}. We find that:
\begin{itemize}
\item they are fully determined by the boundary conditions and the characteristics of their (convective) progenitor (its magnetic- and cross-helicities and its angular momentum that are almost conserved during the relaxation while the energy (magnetic plus kinetic) decays faster following a selective decay \citep[][]{Biskamp1997}).
\item the lowest-energy state for given initial and boundary conditions is the dipolar mode as in the non-rotating case. In the axisymmetric case, the horizontal geometry of the field at stellar surface is dipolar independently of the rotation rate, which modifies only the radial distribution of the magnetic flux. This result is particularly interesting since it matches the observed dipolar geometry of the field both in slowly- and rapidly-rotating early-type stars \citep[][]{Wadeetal2011}. It also points the strong difference between magnetic fields resulting from turbulent relaxation and of those generated by $\alpha-\Omega$ dynamos, which have a geometry that is directly modified by rotation \citep[e.g.][]{Petitetal2008,Brownetal2010}.
\item the field is non force-free \citep[][]{Reisenegger2009,BN2006,DM2010}. However, contrary to the non-rotating case, the configuration is non-torque free because the azimutal component of the Lorentz force must counterbalance advection in the rotating case.
\end{itemize}

\bibliographystyle{iau307}
\bibliography{Biblio_Mathis_proc2}

\end{document}